\documentclass[final,times,twocolumn,5p,a4paper,authoryear]{elsarticle}

\usepackage{amssymb,amsmath}
\usepackage{color}

\usepackage{lineno}

\journal{Journal of Theoretical Biology}

\newcommand{\rev}[1]{\textcolor{black}{#1}}
\newcommand{\revrev}[1]{\textcolor{black}{#1}}

\begin{document}

\begin{frontmatter}



\title{Computation of epidemic final size distributions}


\author[ad1]{Andrew J. Black\corref{cor1}}
\ead{andrew.black@adelaide.edu.au}

\author[ad1]{J. V. Ross}

\cortext[cor1]{Corresponding author: School of Mathematical Sciences, The University of Adelaide, Adelaide SA 5005, Australia. Phone: +61883134177}

\address[ad1]{School of Mathematical Sciences, The University of Adelaide, Adelaide SA 5005, Australia.}

\begin{abstract}

We develop a new methodology for the efficient computation of epidemic final size distributions \rev{for a broad class of Markovian models}. We exploit a particular representation of the stochastic epidemic process to derive a method which is both computationally efficient and numerically stable. The algorithms we present are also physically transparent and so allow us to extend this method from the basic SIR model to a model with a phase-type infectious period and another with waning immunity. The underlying theory is applicable to many Markovian models where we wish to efficiently calculate hitting probabilities.

\end{abstract}

\begin{keyword}
Markov chain \sep degree-of-advancement \sep hitting probabilities \sep waning immunity



\end{keyword}

\end{frontmatter}





\section{Introduction}

Markov chains, in both discrete and continuous time, have found widespread use throughout biology~\citep{Ren91,SLL06,Ros10,Ros11,BM12,HI12}. They are useful as they are amenable to analysis due to the Markovian property but still incorporate the aspect of randomness which is vitally important to accurate modeling of many biological processes \citep{Ren91,BM12}. 
In epidemiology, stochastic models are especially important when considering smaller populations such as households, schools, farms and workplaces \citep{Halloran2008,Keeling:2010,Fraser:2011}. The use of these types of models for inference is becoming more widespread, thus computational efficiency, as well as accuracy, is a primary concern \citep{DON:2006,Brooks:2011,House2012}. For many models, stochastic simulation has been the only way of calculating quantities, but this has the drawback of requiring averaging over many realizations for accuracy. With increasing computing resources, numerical methods of solution, exact to a given precision, are now an attractive proposition for these type of models \citep{Keeling:2009,JG12,BHKR12,BR13}. \\

In epidemiology, one of the most important quantities, for both inference and public health, is the epidemic final size distribution. The final size is the total number of individuals who have experienced infection over the course of the epidemic. \rev{Thus the final size distribution gives the probability of each of the possible outcomes of an epidemic}.  Calculating \rev{the final size distribution} is the subject of a wide body of literature: see \cite{HRS13} for a review. 
For the classic Markovian SIR model, Bailey's \citeyearpar{Bailey:1953,Bai75} method and in particular the implementation due to \cite{NL:1996} has been shown to be superior, being both numerically stable and computationally efficient \citep{HRS13}. \rev{The algorithm} can be derived in a number of ways, but a \rev{clear} physical interpretation has been lacking. \rev{There is also an implicit assumption} that the infectious period is \rev{exponentially distributed which} is known to be unrealistic \citep{Keeling:2007}. Another procedure which allows for more general infectious period distributions is due to \cite{Ball:1986}. Unfortunately, this suffers from a number of numerical problems, even for moderate population sizes \citep{HRS13,DON:2006}.\\

In this paper we present a new method for the computation of final size distributions---\revrev{applicable to homogeneously-mixing Markovian models---}which is both exact and numerically stable. It is also computationally efficient, and physically transparent, allowing us to calculate distributions for a range of more complex models. Our method is based around a particular representation of the stochastic process, know as the {\em degree-of-advancement} \rev{(DA)} representation. This has been recently used in relation to continuous-time Markov chains \citep{Sunkara:2009,JG12}. It is based on counting \rev{events} instead of population numbers and a lexicographical ordering of the state space.\\

The basic idea behind our methodology is intuitively simple. The epidemic process can be considered as a random walk, ending in an absorbing state which then determines the final size. 
The probability of hitting that state is then just the sum of the probabilities of all possible paths of reaching that state. However, it can be difficult to enumerate all these paths and correctly sum them. In fact, in his original paper, \cite{Bailey:1953} considers a suggestion from a colleague for a scheme like this. He rejects the idea though, because `the summation may leave some doubt as to whether all relevant terms have been included'. We show that adopting the DA representation solves this problem and summing over all paths becomes trivial as the jump chain of the process resembles a probability tree. Our method is an advancement on Bailey's method---and derivatives of it---as it is more transparent and can be extended to a range of more complex models; it allows us to calculate final size distributions for models which up until now have been intractable to all but simulation, such as those which include waning immunity. \\

The remainder of this paper is as follows. We begin by illustrating the fundamental idea using  the SIR model. 
We discuss the methodology and algorithms in some detail because the models we consider later are generalizations of this. In particular we compute the final size distribution for an SIR model with a phase-type infectious period distribution and a model with waning immunity where the number of infections is potentially unbounded. Finally we give a discussion of our results and their other uses. In particular, we highlight the connection to the computation of hitting probabilities \citep{Nor97}. Although we have derived these results by considering models in epidemiology, the basic results are much more general. Stochastic models with a similar structure are now common tools in many areas \citep{BM12} and hence this methodology will be potentially useful in a wide range of disciplines. MATLAB code for generating all these results is provided in the supplementary material. \\

\section{SIR final size distribution}

We illustrate the basic idea using the well known Markovian SIR model \citep{Keeling:2007}. The state of the process is $\mathbf{X}(t) = (S,I)$, the number of susceptible and infectious individuals at time $t$. The transitions and rates which define the model are given in Table \ref{tab:SIR}. These transitions are given in terms of the population numbers, $S$ and $I$, hence this is known as the population representation. Instead of this, we work \rev{with the DA} representation \citep{van_kampen,Sunkara:2009,JG12}. This involves counting the number of \rev{events} of each type instead of the population numbers. We therefore define $Z_1$ and $Z_2$ as the number of infection and recovery events respectively, and hence the state of the process at time $t$ is $\mathbf{Z}(t)=(Z_1,Z_2)$. The random variable $Z_1$ also counts the first infection events, which we take to have occurred from an outside source and hence sets the initial condition for the problem. The difference between these representations is that population numbers can both increase and decrease, whereas the transition counts can only ever increase. The number of susceptible, infectious and recovered individuals are then given by, 
\begin{equation}
	S = N-Z_1,  \quad I = Z_1 -Z_2, \quad R = Z_2,
\end{equation}
respectively, where $N$ is the total population size.
Using these relations the rates of each type of transition can be calculated in terms of $Z_1$ and $Z_2$. \\

\begin{table}[ht]
\centering
\begin{tabular}{c | c }
{Transition} & { Rate}  \\ \hline
$ (S,I)\rightarrow (S-1,I+1)$ & $\beta SI$ \\
$ (S,I) \rightarrow (S,I-1)$ & $\gamma I$ \\
\hline
\end{tabular}
\caption{Transitions and rates defining the SIR model. $R=N - S-I$, where $N$ is the size of the population.}
\label{tab:SIR}	
\end{table}

We index the states of the system by $\mathbf{z}_i=(z_1,z_2)$, $i=1,\dots,\Psi$, where $\Psi=(N+1)(N+2)/2$ is the size of the state space. Thus the elements of $\mathbf{z}_i$ are the counts of events $Z_1$ and $Z_2$ that have occurred in reaching the state $i$. The states of the system are ordered, such that $\mathbf{z}_i \prec \mathbf{z}_{i+1}$. This means that,
\begin{equation}
\begin{aligned}
	(z_1,z_2) \prec (z'_1,z'_2) \quad \text{iff}  \quad & z_2<z'_2 \quad \text{or} \\
	& z_2=z'_2 \text{  and   } z_1 \le z'_1.
\end{aligned}
\end{equation}
 This is a {\em co-lexicographic ordering}, in contrast to that used by \cite{JG12}, which is a lexicographic ordering. We choose this because it allows for a simplification of later parts of the algorithm needed to calculate the final size distribution. 
 Figure 1 shows the DA state space for an SIR model with $N=3$. The states are \rev{indexed linearly} according to their co-lexicographic ordering and arrows indicate possible transitions \rev{between states. In practice, this ordering and the linear indexing of the states is most easily enumerated by using a simple nested loop system: the first loop iterates over values of $Z_2=0,\dots,N$ and then a second loop, nested inside the first, iterates over values of $Z_1=z_2,\dots,N$. The state index, $i$, is then simply assigned by keeping count of the number of states which have been iterated over.} 

\begin{figure}[t]
\centering
\includegraphics[width=0.6\columnwidth]{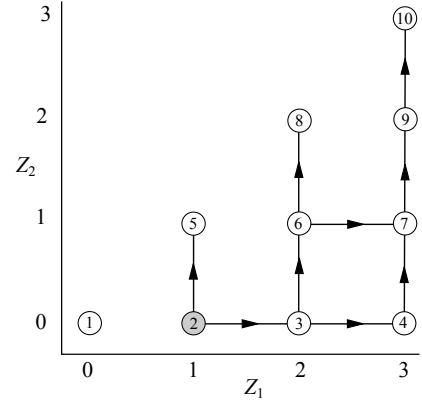}
\caption{The state space of the SIR model in the DA representation. $Z_1$ counts the number of infection events and $Z_2$ counts the number of recovery events. The states are numbered \rev{$i=1,\dots,10$} according to their co-lexicographic order. Arrows denote possible transitions and the initial state, $\mathbf{z}_2=(1,0)$, is colored in gray. The state $\mathbf{z}_1=(0,0)$ corresponds to all individuals being susceptible.}
\label{fig:SIR_DA_ss} 
\end{figure}

This ordering means that the stochastic transition matrix (the generator), and hence the jump chain of the process is upper triangular. The jump chain can now be thought of as a  probability tree, where the probabilities at the leaves (i.e.,~the absorbing states) are found by multiplying and adding probabilities along the different branches. Given that the system starts in state $i$ with probability $b_i$ we can simply write down equations for $p_i$, the probability of visiting (or ending in) state $i$,
\begin{equation}
	p_i = b_i +\sum_{j}\alpha_{ji} p_j
	\label{eq:sum}
\end{equation}
where the sum on $j$ is over all states which lead to state $i$ and $\alpha_{ji}$ is the probability of entering state $i$ from state $j$.  More generally, this system of equations can be written as
\begin{equation}
	(I-A)\mathbf{p} = \mathbf{b},
	\label{eq:DA_solve}
\end{equation}
where $A$ is the transpose of the jump chain matrix. As the matrix $(I-A)$ is lower triangular, this system of equations is solved very efficiently via forward substitution. This is implemented automatically in MATLAB. Once we have solved Eq.~\eqref{eq:DA_solve} then the final size distribution, \rev{$\mathbf{u}=(u_0,\dots,u_N)$}, is found by selecting the elements of $\mathbf{p}$ corresponding to absorbing states, i.e. those where $Z_1=Z_2$. \rev{For this model we can easily write down an expression for indices of these absorbing states,
\begin{equation}
	u_j = p_{j(2N+3-j)/2+1}, \quad j=0,\dots,N.
	\label{eq:abs_map}
\end{equation}}
Solving Eq.~\eqref{eq:DA_solve} gives the probabilities of visiting all states in the system, given that we start from a particular set of states. As the state space for these problems can become very large we might not want to \revrev{store the elements of $A$}. Instead, we can exploit the ordered structure of the state space to derive a simple recursive method for solving equation \eqref{eq:sum}. \\

Firstly we set $\mathbf{p}=\mathbf{b}$, the initial state of the system.
\rev{Next, as described earlier, we use a pair of nested loops to iterate through the states of the system in co-lexicographic order.} For each state we calculate the rates of the two possible events, infection and recovery respectively, as
\begin{equation}
\begin{aligned}
	a_1 =& \beta(N-z_1)(z_1-z_2),\\
	a_2 =& \gamma(z_1-z_2).
	\label{eq:event_probs}
\end{aligned}
\end{equation}
Note that $a_1$ and $a_2$ are \rev{functions of the state}, but we suppress this for clarity of the exposition. We then define the total rate as $a_0=a_1+a_2$.
Remembering that the variable $i$ indexes the states of the system, we then update the elements of $\mathbf{p}$ according to 
\begin{equation}
	\begin{aligned}
		p_{i+1} = p_{i+1} + p_i\, a_1/a_0, \quad a_1 > 0,\\
		p_{i+N-z_2} = p_{i+N-z_2} + p_i\, a_2/a_0, \quad a_2 > 0.
	\end{aligned}
	\label{eq:p_algo}
\end{equation}

This reduces the overhead of the calculation because there is no need to store the elements of the jump matrix and their positions. This algorithm relies on being able to calculate the index of the states which state $i$ feeds into efficiently without the need for complex search routines or pre-calculation of any quantities. The ordering of the state space makes what is potentially a difficult and inefficient procedure quite straightforward. \\

If we do not wish to calculate the whole vector $\mathbf{p}$, but only the probabilities of hitting the absorbing states (i.e.,~the final size distribution) then we can alter the algorithm to require \revrev{even} less storage. This will become important later when we consider problems with a much larger state space. \rev{To see the intuition behind this, firstly note that this algorithm summarised in Eq.~\eqref{eq:p_algo} is different to forward substitution. Each step of the above algorithm  calculates the probabilities of transitioning to the connected states given that we are in state $i$. These parts are then added to any existing transition probabilities which have already been calculated. In this way, each $p_i$ is calculated piecewise in two steps but because of the ordering, once the algorithm reaches state $i$, all the contributions to $p_i$ will have been added and hence $p_i$ can be used to calculate further probabilities. Conversely, forward substitution calculates each $p_i$ in one step from probabilities already calculated as in Eq.~\eqref{eq:sum}. With the algorithm above, once it has iterated over state $i$ then $p_i$ and the reaction rates $a_1$ and $a_2$ are no longer required for the rest of the calculation (unless $i$ is an absorbing state, in which case $p_i$ forms part of the final size distribution). Hence memory which was used to store the probabilities $p_j$, $j\le i$ can be reused. This point underlies all the other algorithms presented in this paper.}\\

\rev{The observation just noted may be used to reduce the amount of memory needed for the algorithm as follows.}
The implementation here assumes that initially the system starts in a state with $m$ infectious individuals. This is the most common situation, but can be extended to handle a situation where there is a distribution of starting states with the addition of some complexity.
\rev{We first define the vector $\mathbf{q}$ which will hold working variables / probabilities. The size of this vector is $|\mathbf{q}| = N+1$ which corresponds to the total number of states for which $Z_2=0$, i.e.,~the bottom row of states in Figure \ref{fig:SIR_DA_ss}.}
Initially the elements of $\mathbf{q}$ are set to zero except \rev{$q_{m+1}=1$}, where $m$ is the initial number of infectious individuals. The  algorithm then proceeds as before, iterating over the states of the system in co-lexicographic order. For each state we calculate the rates as in Eq.\eqref{eq:event_probs} and then update the elements of $\mathbf{q}$ such that,
\rev{
\begin{equation}
\begin{aligned}
	q_{z_1+2} &=q_{z_1+2} + q_{z_1+1}\, a_1/a_0, \quad a_1 >0, \\
    q_{z_1+1} &= q_{z_1+1}\, a_2/a_0, \quad a_2> 0.
    \label{algo:nice}
\end{aligned}
\end{equation}}
\rev{We can see that once the probability stored in element $q_{z_1+1}$ has been used to calculate the two new probabilities, it is replaced with a new probability ($q_{z_1+1}\, a_2/a_0$). Once the algorithm has completed then $\mathbf{q}$ holds the probabilities of the final size distribution ($\mathbf{u}=\mathbf{q}$). Note that for the generalised algorithm in the next section this is not true and the required number of working variables is larger than the final size distribution. }
This version of the algorithm thus reduces the length of the vector needed to store probabilities from $(N+1)(N+2)/2$ to $N+1$, \rev{although the running time is approximately the same as the previous algorithm as we still need to iterate over all the possible states of the system. } \\

As described above, this algorithm is not the most optimized version we can conceive of, but we have left it in this form because it is easier to see how this generalizes to models with many more events. For example, we can optimize this to remove the conditional statements, as we know where the absorbing and boundary states are, but this is more difficult for higher-dimensional models.
The simplicity of the algorithm makes this very computationally efficient. MATLAB code for evaluating this and the previous algorithm is given in the supplementary material.
The implementation of Bailey's method due to \cite{NL:1996} is very similar to the first algorithm \eqref{eq:p_algo} for calculating the full vector $\mathbf{p}$.  Their method uses the population representation of the process hence it is more complicated (MATLAB code for this is presented in the supplementary material of \cite{HRS13}). In contrast, the algorithm above is simpler, \rev{requiring less computational operations and memory. There is also a clear physical interpretation as to the action of the algorithm at each iteration, which is subtly different to that of basic forward substitution.}  As we will now show, it also allows us to efficiently compute final size distributions for much more complicated models.

\section{Model with a phase-type infectious period distribution}

We now consider a model with a phase-type distribution for the infectious period. This is achieved by splitting the infectious period up into $k$ stages, hence these are known as SI($k$)R models. These are widely used in epidemiology as they capture a more realistic infectious profile where \rev{the time an individual remains infectious exhibits less variation about the mean} in comparison to an exponential distribution \citep{Keeling:2007,PN:2009}. We do not consider models with a latent period (such as SEIR) because this has no effect on the final size distribution \citep{Ball:1986}; for time dependent quantities this would not be true.\\

\begin{figure}[!ht]
\centering
\includegraphics[width=0.65\columnwidth]{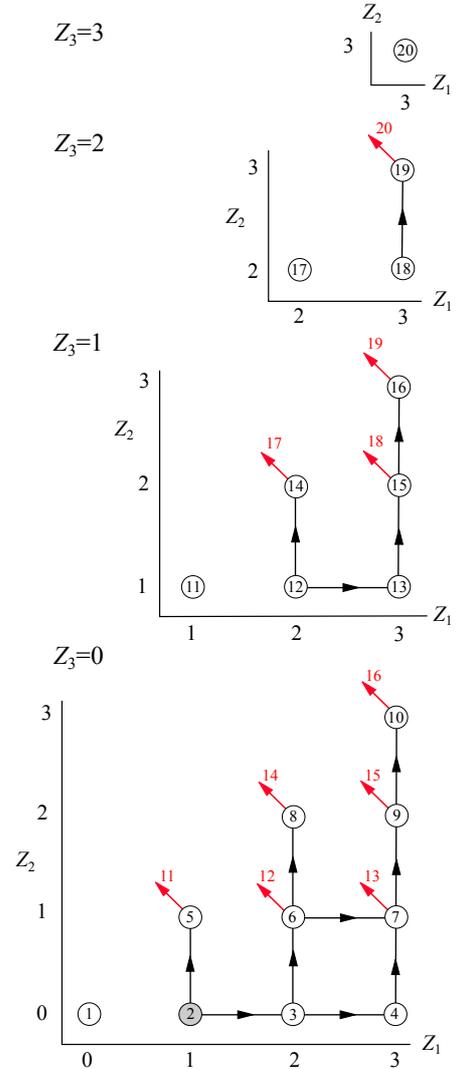}
\caption{\rev{Illustration of the SI(2)R state space with $N=3$ `sliced' by values of $Z_3$. The states are numbered according to their co-lexicographic order. Possible transitions are shown by arrows  with transitions between values of $Z_3$ denoted by red arrows with the destination states also indicated. The algorithm to calculate the final size distribution only requires space to hold $\Phi(3,1)=10$ variables, which corresponds to the number of states with $Z_3=0$, i.e. the bottom slice.}}
\label{fig:si2r} 
\end{figure}

If the infectious period is split into $k$ stages then there will be $k+1$ possible events. We label these from 1 to $k+1$, with the first event being the infection and events 2 through $k+1$ count progress through the infectious stages. \rev{The transition rates of each of the progression events is then $k\gamma$ so that the mean duration of infection stays fixed at $1/\gamma$.} The transitions and rates are summarised in Table \ref{tab:phase}. 
The size of the state space is now,
\begin{equation}
	\Phi(N,k) = \frac{(N+k+1)!}{(k+1)! \,\,N!}=\binom{N+k+1}{N}.
	\label{eq:size}
\end{equation}
\rev{This is the number of ways of allocating $N$ individuals to $k+2$ classes ($S$, $I_1,\dots,I_{k+1}$, $R$).}
\rev{The state space and transitions for this model with $N=3$, $k=2$ are illustrated in Figure \ref{fig:si2r} by slicing the state space according to values of $Z_3$; these should then be visualised as being stacked on top of each other.} \\

The final size distribution can then be found in the same way as for the SIR model, by forming the jump chain for the process and solving equation \eqref{eq:DA_solve}.
An obvious problem with this approach is that with an increasing number of phases, $k$, or population size, $N$, $\Phi$ becomes very large. Thus computing and storing the matrix $A$ becomes expensive, so we need to use a recursive algorithm as discussed in the previous section. \rev{In particular, we do not wish to retain the full vector $\mathbf{p}$, but only wish to  calculate $\mathbf{u}$ using the least memory possible. This reduction in storage is accomplished, as in the previous section, by exploiting the structure of the state space. As can be seen in Figure \ref{fig:si2r}, the slices of the state space for $Z_3>0$ have the same structure as the top part of the $Z_3=0$ slice. Thus we can map the probabilities of reaching these states back into the vector holding the $Z_3=0$ slice probabilities, replacing elements of the vector which are no longer required for the calculation.}\\

\begin{table*}[ht]
\centering	
\begin{tabular}{c | c | c }
{\bf Transition} & {\bf Counter} & {\bf Rate}  \\ \hline
$ (S,I_1)\rightarrow (S-1,I_1+1)$ &  $z_1$  & $\beta (N-z_1)(z_1-z_{k+1})$ \\
\rev{$(I_{j-1},I_{j}) \rightarrow (I_{j-1}-1,I_{j}+1)$} &  \rev{$z_j$ \, $j=2,\dots,k$} & \rev{$k\gamma (z_{j-1}-z_{j})$} \\
$ (I_k,R) \rightarrow (I_k-1,R+1)$ & $z_{k+1}$ & $ k\gamma (z_{k}-z_{k+1})$ \\
\hline
\end{tabular}
\caption{Transitions and rates defining the SI($k$)R model with an Erlang distribution. $R=N-S-\sum_{j=1}^k I_j$, where $N$ is the size of the population. Only states that change in a transition are presented.}
\label{tab:phase}
\end{table*}

\rev{For the SI($k$)R model,} the size of the vector, $\mathbf{q}$, of \rev{working variables} we need to store to compute the final size distribution is $|\mathbf{q}|=\Phi(N,k-1)$, \rev{which corresponds to the number of states of the system with $Z_{k+1}=0$. For the basic SIR model ($k=1$) $\Phi(N,0)=N+1$, which coincides with the number of elements in the final size distribution. For $k>1$ however, the algorithm requires more working variables so the mapping between the states of the system and the elements of $\mathbf{q}$ becomes more complicated. The ratio of the number of states in the system to the number of elements of $\mathbf{q}$ is $(k+1)/(N+k+1)$, so there is an increasing benefit to this formulation of the algorithm as $N$ increases.}\\


\rev{The algorithm proceeds as follows:  Initialization is done by setting } the elements \rev{of $\mathbf{q}$} to zero, apart from $q_{m+1}=1$, where $m$ is the initial number of infectious individuals. As before, we iterate through the states of the system in co-lexicographic order \rev{with the use of a nested loop system. Thus we first loop over values of $Z_{k+1}=0,\dots,N$, which sets the limits for a second loop, nested within the first, which iterates over $Z_k=z_{k+1},\dots,N$, which sets the limits for a third loop {\em et cetera}. The state of the system is then $\mathbf{z}_i=(z_1,\dots,z_{k+1})$, where $i$ is the running total of the  number of states which have been looped over so far. In fact $i$ is not actually needed for any calculations in the algorithm, but is useful for its description.}\\

\rev{We now have to establish a mapping between a given state of the system $\mathbf{z}_i$ and where the probability of reaching that state is stored within the vector $\mathbf{q}$. This is by done by introducing another counter $\omega$ which indexes $\mathbf{q}$. For each new value of $Z_{k+1}$ the algorithm loops over, we calculate
\begin{equation}
	\xi=\Phi(N,k-1)-\Phi(N-z_{k+1},k-1)+1
	\label{eq:offset}
\end{equation}
and then set $\omega=\xi$. After each iteration, $\omega\to\omega+1$, except when $Z_{k+1}$ changes, in which case it is re-calculated from Eq.~\eqref{eq:offset}. So for each value of $Z_3$, $\omega$  counts through the values $\xi,\dots,\Phi(N,k-1)$.}\\

For each value of \rev{$\mathbf{z}_i$} we then calculate the rates of all possible events, $a_j$, $j=1,\dots,k+1$, and their total $a_0=\sum_{j=1}^{k+1} a_j$ and update the elements of $\mathbf{q}$ as follows. For each event type $j=1,\dots,k$ we update\\
\begin{equation}
	q_l = q_l + q_\omega\, a_j/a_0, \quad  a_j >0,
	\label{eq:update1}
\end{equation}
in order, where 
\begin{equation}
	l=\omega+\Phi(N-z_j,j-2).
	\label{eq:l_fun}
\end{equation}
For the \rev{event of type $k+1$, instead, we do}
\begin{equation}
	q_\omega = q_\omega\, a_{k+1}/a_0, \quad a_{k+1} >0.
	\label{eq:update2}
\end{equation}
\rev{These operations are just a generalisation of those given in Eq.~\eqref{algo:nice}, where in the last step the probability stored in $q_\omega$ is replaced by $q_\omega\, a_{k+1}/a_0$, which is the probability of a $Z_{k+1}$ event from the current state.}\\

\rev{Once the algorithm has iterated over all the states of the system, the elements of $\mathbf{u}$ can be extracted from $\mathbf{q}$.} Another way to find $\mathbf{u}$ is to extract the correct probabilities as the algorithm progresses: if it comes to a state which has zero probability of leaving (\rev{i.e.,~when} $Z_{k+1}=Z_1$), then it is an absorbing state corresponding to a final size, $Z_{1}$. Because of the chosen ordering, the order in which absorbing states are encountered matches that of the final size distribution. An example of the output is shown in Figure 3 for the SI($4$)R model and compared with the standard SIR model in the inset. Full code for generating this is given in the supplementary materials.\\

\rev{If $k=1$ then the algorithm presented in this section simplifies as the counter, $\omega$, becomes redundant. This is because, 
\begin{equation*}
\xi=\Phi(N,0)-\Phi(N-z_2,0)+1 = z_2+1,	
\end{equation*}
thus for each value of $Z_2$, $\omega$ would iterate over the values $z_2+1,\dots,N+1$. As $Z_1$ already iterates over $z_2,\dots,N$, then we can simply set $\omega=z_1+1$, and remove the $\omega$ counter. Finally, the function $l$, given by Eq.~\eqref{eq:l_fun}, reduces to $l = z_1+2$ as $\Phi(N-z_1,-1)= 1$. Thus the update rules in Eqs.~\eqref{eq:update1} and \eqref{eq:update2} reduce to those in Eq.~\eqref{algo:nice}.}\\

The power of this recursive approach comes from the simplicity of the algorithm, and hence the speed of execution. Obviously, this relies on being able to calculate the indexes \rev{$\omega$} and $l$ analytically, so this information does not need to be stored. 
Even though the size of the state space for such models can be into the hundreds of millions it can be processed very quickly and numerically it is very stable. This is in contrast to \rev{the method of \cite{Ball:1986}} which, although it uses a much smaller set of equations \revrev{and allows using any distribution for the infectious period}, requires numbers retaining a high degree of accuracy which is harder for a computer to handle natively \citep{DON:2006}.\\

\begin{figure}[t]
\centering
\includegraphics[width=0.8\columnwidth]{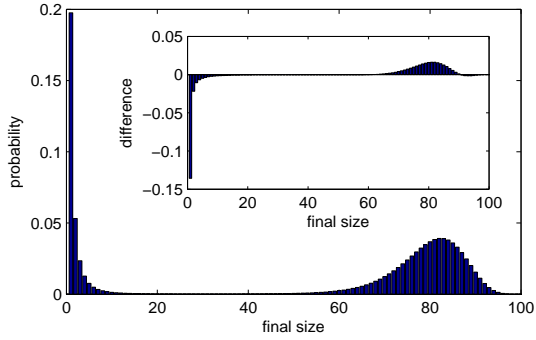}
\caption{Main plot shows the \rev{probability mass function} for a SI(4)R model, with $N=100$. The inset shows the difference between the pmfs for the Erlang(4) model and the basic exponential SIR model. Both epidemics start with 1 infectious individual. Other parameters: $\beta=2/(N-1)$ and $\gamma=1$.}
\label{fig:phase} 
\end{figure}

\begin{figure*}[t]
\centering
\includegraphics[width=1.8\columnwidth]{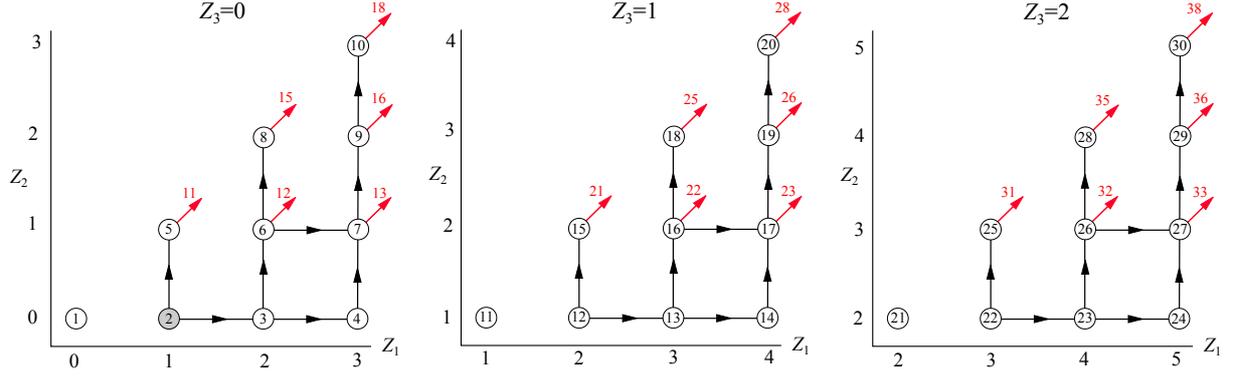}
\caption{Illustration of the structure of state space for the SIRS model with $N=3$ only showing the first three slices for $Z_3=0,1,2$. States are labelled by their co-lexicographic order. For each value of $Z_3$ the slice of state space \rev{has the same structure as} the triangular state space of the SIR model in Figure 1. Possible transitions between slices of constant $Z_3$ are shown with the red arrows along with the destination state.}
\label{fig:SIRS_ss} 
\end{figure*}

\section{Model with waning immunity}

We now use the methodology developed in the previous sections to calculate the total number of infections in a stochastic epidemic model with waning immunity. The simplest such model is a \emph{susceptible-infected-recovered-susceptible} (SIRS) model \citep{Keeling:2007}. The transitions defining this are as in Table \ref{tab:SIR} with the addition of 
\begin{equation}
	(S,R) \rightarrow (S+1,R-1) \quad \text{at rate}\quad \mu R,
\end{equation}
where $R=N-S-I$ is the number of recovered individuals. 
\rev{The state space in the population representation now contains loops and there is only one absorbing state corresponding to $(S=N, I=0, R=0)$. Thus, in this representation, we can no longer calculate the final size distribution (defined as the total number of infection events) from hitting probabilities. To move to the DA representation, we first}
denote the number of waning immunity events by $Z_3$. Such a model is different to the SIR model \rev{presented in Section 2} as there is no upper bound on the total number of infection events. Thus,
\begin{equation}
	S = N-Z_1+Z_3, \quad I = Z_1-Z_2, \quad R = Z_2-Z_3,
	\label{eq:SIRS_pops}
\end{equation}
and the state space is unbounded. 

\rev{The first part of the state space of this model,  with $N=3$, is illustrated in Figure \ref{fig:SIRS_ss}. Each slice (constant value of $Z_3$) has the same structure as the SIR model shown in Figure \ref{fig:SIR_DA_ss}. The red arrows indicate $Z_3$ transitions between different slices. Importantly, there are no loops in this state space  because the counts can only ever increase. Thus the absorbing states are when $Z_1=Z_2=Z_3$, which corresponds to a final size of $Z_1$.} \\

One way \rev{to calculate the final size distribution} is to simply truncate the DA state space, form the jump chain and solve Eq.~\eqref{eq:DA_solve}. A much better way is to exploit the ordered structure of the state space to create a recursive method. By ordering the states with respect to the number of $Z_3$ events, as in Figure \ref{fig:SIRS_ss}, we can use Eqs.~\eqref{eq:SIRS_pops} to derive limits on the possible number of $Z_1$ and $Z_2$ events,
\begin{equation}
	Z_3 \le Z_1 \le N+Z_3, \quad\text{and}\quad Z_3 \le Z_2 \le N+Z_3.
\end{equation}
This suggests we make a transformation of variables, $Z_1'=Z_1-Z_3$ and $Z_2'=Z_2-Z_3$, so the rates of events $Z_1$, $Z_2$ and $Z_3$ become,
\begin{equation}
	\begin{aligned}
		a_1 &= \beta(N-z_1')(z_1'-z_2'),\\
		a_2 &= \gamma (z_1'-z_2'),\\
		a_3 &= \mu z_2',
	\end{aligned}
	\label{eq:SIRS_rates}
\end{equation}
respectively. Note, that these rates are now independent of $z_3$.\\

To recursively compute the final size distribution (total number of $Z_1$ events) we only need to store probabilities for \rev{the number of states in} two slices of the state space. \rev{We denote these working variables as $\mathbf{q}$ and $\mathbf{h}$}, both of length $\Psi=(N+1)(N+2)/2$. We truncate the maximum number of $Z_1$ events, and hence the largest possible observed final size \revrev{at} $\Lambda$. We then define the vector $\mathbf{u}=(u_0,\dots,u_{\Lambda+1})$ which will hold the final size distribution. The last element of $\mathbf{u}$ will be probability of observing more than $\Lambda$ infection events, which is also computed very naturally using this approach.

 We initialize \rev{$\mathbf{q}$}, $\mathbf{h}$ and $\mathbf{u}$ by setting all their elements to zero, apart from $q_{m+1}=1$, where $m$ is the initial number of infectious individuals. The algorithm then proceeds as follows: for each value $Z_3=0,\dots,\Lambda$, we iterate over all possible values of $Z_1'$ and $Z_2'$ in co-lexicographic order in exactly the same way as for the basic SIR model. \rev{As in the previous algorithm, we maintain a counting variable, $\omega$, which counts the number of states that have been iterated over for a given value of $Z_3$. The maximum value of $\omega$ will be $\Psi$ as this is the number of states in one slice of the state space.}\\

 For each state we calculate the rates of the three events from Eq.~\eqref{eq:SIRS_rates}. We then update the elements of $\mathbf{q}$ and $\mathbf{h}$ as follows:
\rev{\begin{equation} 
	\begin{aligned}
		q_{\omega+1} &= q_{\omega+1} + q_\omega\, a_1 /a_0, \quad a_1 > 0, \\
		q_{\omega+N-z_2'} &= q_{\omega+N-z_2'} + q_\omega\, a_2/a_0, \quad a_2 > 0, \\
		h_{\omega-N+z_2'-2} &= q_\omega \, a_3/a_0, \quad a_3 > 0,
		\end{aligned}
	\label{eq:SIRS_algo}	
\end{equation}}
where $a_0=a_1+a_2+a_3$. \rev{The first two equations in \eqref{eq:SIRS_algo} have the same form as for the SIR model given in Eq.~\eqref{eq:p_algo}.}
In determining the element of $\mathbf{h}$ to update we have taken into account the transformation of variables above. \rev{Once the algorithm has iterated over all values of $Z_1'$ and $Z_2'$ we can update $\mathbf{u}$ as,
\begin{equation}
	u_{z_3} = q_1.
\end{equation}}
\rev{This follows because, for a given value of $Z_3$, the absorbing state is when $Z'_1=Z'_2=0$ which is stored in the first element of $\mathbf{q}$.} The algorithm then continues by setting $\mathbf{q}=\mathbf{h}$, then $\mathbf{h}=0$ and looping to the next value of $Z_3$. Once all values of $Z_3$ have been iterated over, the probability of more than $\Lambda$ infection events is simply
\begin{equation}
	u_{\Lambda+1} = \sum_{j=1}^\Psi q_j.
\end{equation}
\rev{This sum is over the elements of $\mathbf{q}$ because in the last step of the algorithm the probabilities stored in $\mathbf{h}$ are moved to $\mathbf{q}$. Alternatively, $u_{\Lambda+1}$ could be calculated as $1- \sum_{j=1}^\Lambda u_j$.} \rev{Because of the ease with which we can calculate this probability, an alternative way to formulate the algorithm is to only terminate the recursion over $Z_3$ once the probability of more than a given number of infections falls below some pre-defined threshold.} Figure 5 shows the probability mass function of the total number of infections for an SIRS model with $N=30$.

\begin{figure}[t]
\centering
\includegraphics[width=0.8\columnwidth]{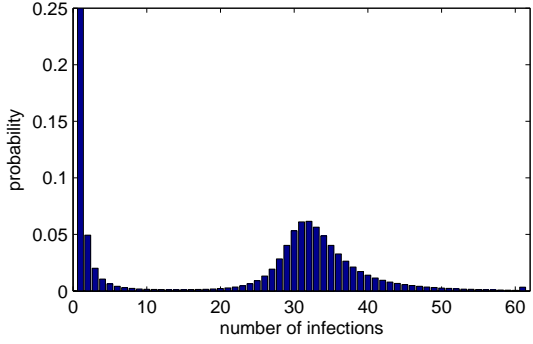}
\caption{The probability mass function, $\mathbf{u}$, for the total number of infections in an SIRS model with $N=30$. The maximum number of infections, \rev{$\Lambda$}, is set to 60. The bar at 61 is then the probability of $Z_1 > 60$. Parameters are: $\beta=3/(N-1)$, $\gamma=1$ and $\mu=0.1$. }
\label{fig:SIRS_pmf1} 
\end{figure}

\section{Discussion and Conclusion}

We have presented a new method for computing the final size distribution for a number of epidemic models. Using the DA representation of the stochastic process means the jump chain has a particularly simple form resembling a probability tree. This structure allows the derivation of an iterative method for calculating the final size distribution. Although similar methods have been proposed for the SIR model \citep{NL:1996}, ours is both the simplest to implement and understand and is also more efficient computationally and in terms of memory usage. The biggest advantage of our method is its straightforward applicability to more complex models. We have demonstrated this by computing final size distributions for the SI($k$)R model and the SIRS model, where the number of infection events is unbounded.  This opens up these models for use in inference work using final size data \citep{HRS13}.\\

Although we have derived and illustrated these methods using models within epidemiology, they are much more general. They are applicable to Markovian models in which we wish to calculate hitting probabilities \citep{macnamara_stochastic_2010} or where probabilities of given paths through a state space are required \citep{nowak_mechanisms_2009,Williams:2013}.  In ecology and epidemiology we are often interested in offspring distributions. This is the distribution of the number of secondary entities a single entity gives rise to over a particular lifetime. \cite{Ros11} has shown how offspring distributions for a stochastic model can be computed by recursively solving sets of linear equations. With the methodology presented here, these quantities could be computed more efficiently in much the same way as presented for the SIRS model.\\

A limitation on this method is that as more event types are added the size of the state space grows as described by Eq.~\eqref{eq:size}. Thus although typically we are not limited by memory due to the recursive nature of the solutions, the time to compute a distribution will grown linearly with the size of the state space. As an example of times for computation, the SI(4)R model shown in Figure 3 has $N=100$ and $\Phi\approx4.6\times 10^6$; computation of the final size distribution is approximately 3.1 seconds using an algorithm programmed in C on a 2.5GHz machine. The same model with $N=200$ has $\Phi\approx 70\times 10^6$; this took 1.8 minutes to process. By normal standards, these are very large state spaces. One advantage of the Ball method is that it can handle any infectious period distribution for which we can write down the Laplace transform \citep{Ball:1986}, whereas our method is limited to phase-type distributions.\\

For very large values of $k$, the number of stages in the infectious period, Ball's method combined with a variable precision arithmetic \citep{DON:2006} might outperform our method, depending on the implementation of the variable precision arithmetic. However, for values of $k$ typically of interest, our method is the most efficient. \rev{Another important class of epidemic models are those with a structured population such as household or network models \citep{Ball:1997,Danon:2011}. These have such large state spaces, even for modest population sizes, that our method would become impractical.} \revrev{For such models a number of good techniques have been developed \citep{ONeill:2009,Neal:2012}.}  \\

Fundamentally the algorithm we have presented is a serial process, so there is little possibility of using parallel computational resources to speed up the procedure.
One promising direction which could allow even larger models is the use of an adaptive mechanism which limits the algorithm to the parts of the state space where most probability is concentrated. The ordering of the state space lends itself to a scheme like this so, especially as the dimensionality of the problem increases, this could provide large savings with little loss of accuracy. For larger values of $N$ there is also the possibility of deriving approximate analytic solutions \rev{from asymptotic results}. \\

There are two other areas where this methodology is potentially important. The DA representation was originally conceived as a way of speeding up the integration of the forward equation which describes the temporal dynamics of the system \citep{JG12}. This method becomes slow as the size of the state space increases, thus it would be very beneficial to be able to restrict the state space to a smaller region as in finite-state projection methods \citep{Munsky:2006}. Our methodology provides a way to calculate which regions of the state space should be kept by solving for the vector $\mathbf{p}$ which provides information about the most probable paths through the system. Another potentially important use of this methodology is in computing the probability of hitting a {\em fixed} state from a given set of initial states \citep{Nor97}. We often want to calculate these quantities when working with conditioned Markov chains \citep{Waugh:1958}. This quantity can be computed in a very similar way to the final size distribution except that the iteration over the states of the system proceeds backwards from the final state.  Again the structure of the state space makes this straightforward to implement. \\

\section*{Acknowledgements}

This research was supported under the Australian Research Council's Discovery Project (DP110102893) and Future Fellowship (FT130100254) funding schemes. We would also like to thank the two anonymous referees whose extensive comments and careful reading have improved this paper. 

\section*{References}


\newcommand{\SortNoop}[1]{}





\end{document}